\documentclass[sigconf]{acmart}

\renewcommand\footnotetextcopyrightpermission[1]{} 
\usepackage{algorithm}
\usepackage{algpseudocode}
\usepackage{svg}
\usepackage{graphicx}
\newcommand{\indep}{\rotatebox[origin=c]{90}{$\models$}}

\AtBeginDocument{%
  \providecommand\BibTeX{{%
    \normalfont B\kern-0.5em{\scshape i\kern-0.25em b}\kern-0.8em\TeX}}}
    
\setcopyright{acmcopyright}
\copyrightyear{2022}
\acmYear{2022}
\acmDOI{XXXXXXX.XXXXXXX}

\acmPrice{15.00}
\acmISBN{978-1-4503-XXXX-X/18/06}

\author{Junliang Luo}
\affiliation{%
  \institution{McGill University}
  \city{Montréal, Québec}
  \country{Canada}
}
\email{junliang.luo@mail.mcgill.ca}

\author{Yongzheng Jia}
\affiliation{%
  \institution{UC Berkeley}
  \city{Berkeley, California}
  \country{USA}}
\email{abner@berkeley.edu}

\author{Xue Liu}
\affiliation{%
  \institution{McGill University}
  \city{Montréal, Québec}
  \country{Canada}
}
\email{xueliu@cs.mcgill.ca}

\setcopyright{none}

\begin{document}


\title{\huge Understanding NFT Price Moves through Tweets Keywords Analysis}



\renewcommand{\shortauthors}{}

\begin{abstract}
Non-Fungible Token (NFT) is evolving with the rise of the cryptocurrency market and the development of blockchain techniques, which leads to an emerging NFT market that has become prosperous rapidly then followed by a cooldown.
Nevertheless, the overall rise procedure of the NFT market has not been well understood.
To this end, we consider that social media communities evolving alongside the market growth, are worth exploring and reasoning about, as the mineable information might unveil the market behaviors.
We explore the procedure from the perspective of NFT Twitter communities and its impact on the NFT price moves with two experiments.
We perform a Granger causality test on the number of tweets and the NFT price time series and find that the number of tweets has a positive impact on (Granger-causes) the price or reversely for larger part of the 19 top authentic projects but seldom copycat projects.
Besides, to investigate the price moves predictability, we experiment on predicting Markov normalized NFT price (representing the direction and magnitude of price moves) given tweets-extracted word features and interpret the feature importance to find insights.
Our results show that social media words as the predictors result in all 19 top projects having a testing accuracy evidently above the baseline.
Based on the feature importance analysis, we find that both general market-related words and NFT event-related words have a markedly positive contribution in predicting price moves. 
We summarize the characteristics including categorization and sentiment for the words with the top and least feature importance.
\end{abstract}






\maketitle
\pagestyle{plain}

\begin{table*}[!htbp]
\huge
\caption{NFT collections of tweets and NFT transaction data from Sept 2018 to Nov 15, 2022}\label{tab:total_table}
\centering
\resizebox{0.89\textwidth}{!}{%
\begin{tabular}{@{}lllllll@{}}
\toprule
Project (Collection) & Assets & Contract Address  & Twitter Account & Originality & Tweets (after filtering) & Transactions \\ \midrule
CryptoPunks & 10K \quad\quad\quad\quad\quad\quad & 0xb47e3cd837ddf8e4c57f05d70ab865de6e193bbb \quad\quad\quad\quad\quad\quad\quad & @cryptopunksnfts & orig.  & 8877 & 49052 \\
Bored Ape Yacht Club & 10K & 0xbc4ca0eda7647a8ab7c2061c2e118a18a936f13d & @BoredApeYC & orig. & 323317 & 91767 \\ 
Mutant Ape Yacht Club \quad\quad\quad\quad\quad\quad & 20K & 0x60e4d786628fea6478f785a6d7e704777c86a7c6 & @BoredApeYC & orig. & 323317 & 120323 \\
Otherdeed for Otherside & 100K & 0x34d85c9cdeb23fa97cb08333b511ac86e1c4e258 & @OthersideMeta & orig. & 36989 & 280441 \\
Art Blocks Curated & 56K & 0xa7d8d9ef8d8ce8992df33d8b8cf4aebabd5bd270 & @artblocks\_io & orig. & 26584 & 516227 \\
Azuki & 10K & 0xed5af388653567af2f388e6224dc7c4b3241c544 & @AzukiOfficial & orig. & 52551 & 74822 \\
CLONE X & 20K & 0x49cf6f5d44e70224e2e23fdcdd2c053f30ada28b & @RTFKTstudios & orig. & 54622 & 83920 \\
Decentraland & 97.5K & 0xf87e31492faf9a91b02ee0deaad50d51d56d5d4d & @decentraland & orig. & 51692 & 209737 \\
The Sandbox & 158K & 0x5cc5b05a8a13e3fbdb0bb9fccd98d38e50f90c38 & @TheSandboxGame & orig. & 62892 & 199681 \\
Moonbirds & 10K & 0x23581767a106ae21c074b2276d25e5c3e136a68b & @moonbirds & orig. & 29588 & 37244 \\
Doodles & 10K & 0x8a90cab2b38dba80c64b7734e58ee1db38b8992e & @doodles & orig. & 90020 & 77811 \\
Meebits & 20K & 0x7bd29408f11d2bfc23c34f18275bbf23bb716bc7 & @MeebitsNFTs & orig. & 3418 & 81230 \\
Cool Cats & 10K & 0x1a92f7381b9f03921564a437210bb9396471050c & @coolcatsnft & orig. & 74437 & 65500 \\
Bored Ape Kennel Club & 9.6K & 0xba30e5f9bb24caa003e9f2f0497ad287fdf95623 & @BoredApeYC & orig. & 323317 & 58787 \\
Loot (for Adventurers) & 7.8K & 0xff9c1b15b16263c61d017ee9f65c50e4ae0113d7 & @lootproject & orig. & 4389 & 31328 \\
CryptoKitties & 2M & 0x06012c8cf97bead5deae237070f9587f8e7a266d & @CryptoKitties & orig. & 10883 & 3355686 \\
CrypToadz & 7K & 0x1cb1a5e65610aeff2551a50f76a87a7d3fb649c6 & @cryptoadzNFT & orig. & 21089 & 44031 \\
World of Women & 10.0K & 0xe785e82358879f061bc3dcac6f0444462d4b5330 & @worldofwomennft & orig. & 62011 & 52127 \\
SuperRare & 32K & 0xb932a70a57673d89f4acffbe830e8ed7f75fb9e0 & @SuperRare & orig. & 111295 & 96777 \\
Phunky Ape Yacht Club & 10K & 0xa4009d8eda6f40f549dfc10f33f56619b9754c90 & @phunkyApeYC & fake BAYC & 9683 & 3418 \\
PHAYC & 10K & 0xcb88735a1eae17ff2a2abaec1ba03d877f4bc055 & @phaycbot & fake BAYC & 2231 & 14687 \\
CryptoPhunks & 10K & 0xf07468ead8cf26c752c676e43c814fee9c8cf402 & @CryptoPhunksV2 & fake CryptoPunks & 37675 & 37917 \\
SameToadz & 7K & 0x07e5ce0f8fa46031a1dcc8cb2530f0a52019830d & @SameToadz & fake CrypToadz & 1211 & 13089 \\
AIMoonbirds (AINightbirds) & 10K & 0x64b6b4142d4d78e49d53430c1d3939f2317f9085 & @ainightbirds & Moonbrid copycat & 21456 & 51971 \\
Zukibirds & 4.1K & 0x0d1fe1ebab085bd039b4d1fbf96dbe8decf769a1 & @Zukibirds & Azuki, Moonbirds copycat & 3175 & 12876 \\
MoonbirdPunks & 2.5K & 0x266a5797e803e5e299a806afa51fb2d80ec31911 & @moonbirdpunks & CryptoPunks, Moonbirds copycat \quad\quad\quad\quad\quad\quad & 801 & 5839  \\
Undead Pastel Club & 10K & 0x0811f26c17284b6e331beaa2328471107576e601 & @UndeadPastelNFT & BAYC copycat & 9849 & 27226 \\
Coodles & 8.9K & 0x9c38bc76f282eb881a387c04fb67e9fc60aecf78 & @CoodlesNFT & Doodle copycat & 5201 & 25705 \\
Lil Baby Ape Club & 5K & 0x918f677b3ab4b9290ca96a95430fd228b2d84817 & @LilBabyApeClub & BAYC copycat & 10840 & 16350 \\
Lil Baby Cool Cats & 5K & 0x838cd6b5bf716ecb1529670850b7265a2d1bbd7c & @LilBabyCoolCats (no acct.) \quad\quad\quad & Cool Cats copycat & 2053 & 8141 \\ \bottomrule
\end{tabular}
}
\label{NFT_collection_table}
\end{table*}

\section{Introduction}
Non-Fungible Token (NFT), with the earliest recognized example created in 2014 by McCoy and Dash \cite{dash2021nfts} registering a video clip on a blockchain, refers to a type of blockchain token that is unique and non-replicable so that it can designate the ownership of artwork, in-game items, domain names, assets in decentralized finance (DeFi), etc. \cite{mazur2021non}.
The ownership is stored in a decentralized manner utilizing the emerging blockchain technique, which deters any centralized authority from owning the distinct right to remove the ownership.
After the early pioneer stage, the NFT standard in Ethereum ERC721 \cite{williamentriken2018} introduced in 2017 shaped the mainstream NFTs projects by standardizing NFT creation, transfer, and project deployment.
The standardization is followed by an emerging of NFT marketplaces to trade NFTs.
NFT marketplaces have their users trading NFTs of art images, music, gaming cards, domain names, etc., for cryptocurrency.
Opensea, as the largest NFT marketplace, has reached a total record of \$31 billion volume and 1.8 million traders as of June 2022 \cite{dappradar}.
The NFT trading volume as of May 2022 exceeds \$37 billion, close to the total of \$40 billion in 2021 \cite{theblock_2022} even though the NFT market after 2022 April has been accompanied by a bearish market of cryptocurrency and NFT in the second half of 2022.
As the NFTs trading volume used to rise for two years, the data generated by the market began to unveil what in practice NFT contributes as an innovation.
The close relation between NFTs and artwork will probably bring to mind that NFTs help encourage art creativity by making small individual artists connect with wider collectors to earn more profits.
However, recent research by Vasan et al. \cite{vasan2022quantifying} presents that despite significantly more artists joined in NFT digital art market, the artist clusters are driven by homophily, i.e., successful artists invite successful artists into the NFT market and create similar sales patterns.
They highlight the forming of the artist–collector ties, i.e. some successful artists receive repeated investment from a small group of collectors. 
According to Nadini et al \cite{nadini2021mapping}, as of April 2021, the top 10\% of buyers-sellers pairs contributed to the number of transactions the same as the rest 90\%.
These findings make us consider what NFT ecosystem builds may be pertaining to the members tie and communities.
The buyers willing to buy an expensive NFT from a collection expect the NFT to be an entrance ticket to join the community behind the collection.
%
The possession of those NFTs may have brought some private connections to the owners as well as some member benefits such as the right to receive airdrops of new related project NFTs, or to join in making decisions on the project's funds, etc.
Apart from the private connections and benifits, the project teams also build public social media communities mostly on Twitter or Discord.
The community interaction about the development of the project persists to attract not just the NFT owners but every user on social media to join and engage in the communities.
We consider that the rise of the prices would be witnessed by the communities formed on social media, so we intend to investigate the relationship between social media content and the NFT price moves to understand the NFT market.
In our work, we collect the historical tweets and NFT trade transactions of 19 top collections ranked by Opensea in volume and 11 corresponding copycat collections (for RQ2).
We conduct research for the following research questions:
RQ1: Does the activeness of a social media community help in forecasting the prices and or reversely? 
RQ2: Will the causal relationship be inspected stronger in authentic NFT projects than the copycat projects since copycat project may not have sustained social media community building as same as the authentic projects?
RQ3: Are the social media word features good predictors for the direction and magnitude of NFT price moves? More importantly, which word features mostly affect the price moves?
To answer RQ1, we perform a Granger causality test on the time series data of the number of tweets and the average trading price for each project. 
For RQ2, we compare the Granger causality test results of the authentic collections with the results of the copycat collections. 
%
%
For answering RQ3, we set up a prediction task where first we divide both tweets and transactions into segments by timeframes.
We apply a words vector extraction method based on term frequency-inverse document frequency (TF-IDF) \cite{jones1972statistical} to extract the important words for each timeframe by treating the words in each timeframe as a document. 
%
%
We have the TF-IDF scores for each word to be used as the features along with a Markov normalized average NFT trading price to be used as the ground truth for each timeframe.
We add lags to the words features for predicting the future NFT price given the past words features.
We experiment on three regression models: SVM, MLP, and Transformer to perform a prediction task and analyze the feature importance.
We summarize the contributions: 

\begin{itemize}
     \item  We investigate the causality relation between the activeness of a Twitter community and NFT price on all the projects. Our results show that 10 out of 19 authentic projects show marked Granger causality between the number of tweets and NFT price or reversely, while only 2 out of 11 copycat projects show Granger causality. 
    The results evidence the NFT price or the Twitter community activeness is useful in forecasting the other for a project and support our hypothesis that copycat projects show weaker evidence. 
    \item
    We explore the NFT price predictability given social media word features with three machine learning models.
    Our empirical results show that all 19 authentic projects have a testing accuracy evidently above the random baseline showing a degree of predictability and the Transformer model performs better than the other models.
    \item
    We analyze the feature importance of the word features and find that the general market-related words and the NFT event-related words appear more in words with the top positive importance values than those with negative importance values, which means that the two types of words contribute more in predicting the NFT price. 
    We adopt a pretrained zero-shot classification and a pretrained sentiment inference model for categorizing and extracting the sentiment of the words. We found that the category of a word (market, NFT event, or regular community) is more relevant to whether the word is a good predictor for price moves compared to sentiment. 
    The results model which produces feature importance, categorization, and sentiment analysis based on price moves prediction given keywords can be used to lay a foundation for further development of analytics to understand the NFT community and market.
\end{itemize}

\section{Related work}
Recent research related to our work includes NFT market analysis and NFT social media studies.
Other related works concern the social media text study for stock movement prediction \cite{sawhney-etal-2020-deep}, for crude oil market price prediction \cite{elshendy2018using}, and for global cryptocurrency price trend prediction \cite{poongodi2021global}.
%


\subsection{NFT Market Analysis}
NFT market has had tremendous growth in trading volume over the past two years.
Nadini et al. characterized the market statistical properties such as the distribution of average price and sales per NFT from June 2017 to April 2021 \cite{nadini2021mapping}.
They also investigated the predictability of NFT sales given the sale history and NFT price given the visual features.
White et al. analyzed the sales data from OpenSea between Jan 2019 and Dec 2021 and found that a small group of whale NFT collectors are driving massive market growth \cite{white2022characterizing}. 
Franceschet proposed a rating method for utilizing artists and collectors trading networks and evaluates the data of the SuperRare NFT market, then has some network metrics to suggest investment strategies \cite{franceschet2021hits}.
Besides, other research discussed the potential fraudulent behaviors in the NFT emerging market, recent studies summarized malicious behaviors in NFT space \cite{rehman2021nfts, kshetri2022scams}. 
Das et al. performed an analysis on NFT marketplaces to discuss the security issues the NFT market is facing and one of the issues is the Counterfeit NFT creation \cite{das2022understanding}.
They performed the quantitative analysis on counterfeit NFTs created by searching all NFTs in markets to find counterfeit NFT collections with similar collection names, identical image URLs, or similar images as some authentic NFTs.
We note the lack of research on comparing the authentic NFT projects and the counterfeit NFT projects.
To this end, we begin to study from the social media perspective, investigate the relation between social media activeness and NFT price and compare the authentic NFT projects with the counterfeit NFT projects.
\subsection{NFT Market and Social Media}
Before, academic work has been conducted on the interaction between social media and cryptocurrency.
One such example is the social media indicator for cryptocurrency price moves prediction \cite{ortu2022technical}.
Besides, Phillips et al. investigated which certain topics discussed on social media are indicative of cryptocurrency price moves using a statistical Hawkes model \cite{hawkes1971spectra} and they illustrated the results by the words that precede positive or negative return \cite{phillips2018mutual}.
Also Mendoza-Tello et al. analyzed the impact of social media on increasing the trust to use cryptocurrencies \cite{mendoza2018social}.
In addition, Nizzoli et al. studied the social media manipulation patterns.
They detected social media bot accounts that broadcast suspicious links and summarized the deception schemes in online cryptocurrency communities \cite{nizzoli2020charting}.

NFT is traded on cryptocurrency marketplaces and it shares a tight relation with cryptocurrency technically.
NFT projects also evolve with the NFT communities emerging similar to cryptocurrency.
Social media plays an essential role in NFT community development since platforms such as Twitter, Reddit, or Discord become where people know about new events for the NFT projects.
Recent works showed that the social media features make improvements for an NFT valuation classification task \cite{kapoor2022tweetboost}.
Aside from the market, Casale-Brunet et al. analyzed the NFT communities on Twitter using social network analysis \cite{casale2022impact}.
They found that most top NFTs can be considered as a single community, where most top projects are influenced by the development of the Bored Ape Yacht Club\footnote{https://boredapeyachtclub.com} collection from a social network perspective.
However, we note the lack of a study investigating the relationship between the language content in these communities to NFT price growth.
Therefore, we seek the extension of the analysis for the impact of important words used in social media communities on NFT price moves in our work.

\section{Data Collection}
\label{data_collection}
We collect the NFT token trade transactions for some most successful NFT collections from Opensea top 19 (top 20 exclude marketplace Rarible) as of 2022 first half year  and their corresponding fake or copycat collections in Table. \ref{NFT_collection_table}.
The data was collected by querying Google BigQuery\footnote{https://cloud.google.com/bigquery} \textit{bigquery--public--data.crypto\_ethereum} resource given the smart contract addresses of those NFT collections.
The transactions contain \textit{address from}, \textit{address to}, \textit{token id}, \textit{transaction hash}, \textit{transaction value} (price in Eth wei), \textit{transaction hash}, and \textit{block timeframe}.
The transactions are either with a transaction value of 0 or a small positive value as a transfer or with a positive transaction value as a sale emitted by the smart contract.
If more than one NFTs were exchanged in the same transaction, we split the transaction value equally.
We collect all the tweets from the account, all the replies to these tweets, and the other tweets @account name or mention the account name for each Twitter account in Table. \ref{NFT_collection_table}, corresponding to the period between Jan 01, 2018 to Nov 15, 2022, except for CryptoKitties from Sept 16, 2018 to Nov 15, 2022 due to the large volume.
We filter out the tweets with less than five likes for the original collection and less than one likes for copycat projects to drop some noisy data.

\begin{table}[!htbp]
\huge
\caption{Squared Residuals (SSR)-based F Granger causality test. Null Hypothesis (NH): A: The number of tweets does not Granger cause the price of NFTs. B: The price of NFTs does not Granger cause the number of tweets. Bold P-value indicates that the null hypothesis of no Granger causality can be rejected at the 0.05 level. A midrule line separate the original and the copycat projects.}\label{tab:total_table}
\centering
\resizebox{0.49\textwidth}{!}{%
\begin{tabular}{@{}lllccccccccccccc@{}}
\toprule
Project (Collection) & NH &  & \multicolumn{1}{l}{F-statistic} & \multicolumn{1}{l}{P-value} & \multicolumn{1}{l}{Corr P-value} & \multicolumn{1}{l}{} & \multicolumn{1}{l}{} & \multicolumn{1}{l}{F-statistic} & \multicolumn{1}{l}{P-value} & \multicolumn{1}{l}{Corr P-value} & \multicolumn{1}{l}{} & \multicolumn{1}{l}{} & \multicolumn{1}{l}{F-statistic} & \multicolumn{1}{l}{P-value} & \multicolumn{1}{l}{Corr P-value} \\ \midrule
 &  &  & \multicolumn{3}{c}{Lags: 1 tf} & \multicolumn{1}{l}{} & \multicolumn{1}{l}{} & \multicolumn{3}{c}{Lags: 2 tfs} &  &  & \multicolumn{3}{c}{Lags: 3 tfs} \\ \cmidrule(lr){4-6} \cmidrule(lr){9-11} \cmidrule(l){14-16} 
CryptoPunks & A &  & 9.354 & \textbf{0.002} & \textbf{0.01} &  &  & 4.725 & \textbf{0.009} & \textbf{0.045} &  &  & 2.871 & \textbf{0.035} & 0.131 \\
 & B &  & 13.441 & \textbf{0.0003} & \textbf{0.002} &  &  & 7.348 & \textbf{0.0007} & \textbf{0.007} &  &  & 5.474 & \textbf{0.001} & \textbf{0.015} \\
Bored Ape Yacht Club & A &  & 5.687 & \textbf{0.018} & 0.077 &  &  & 3.872 & \textbf{0.022} & 0.082 &  &  & 3.569 & \textbf{0.015} & 0.075 \\
 & B &  & 12.420 & \textbf{0.0005} & \textbf{0.002} &  &  & 5.082 & \textbf{0.007} & 0.052 &  &  & 2.699 & \textbf{0.047} & 0.235 \\
Mutant Ape Yacht Club & A &  & 0.426 & 0.514 & 0.811 &  &  & 1.662 & 0.193 & 0.445 &  &  & 1.518 & 0.212 & 0.530 \\
 & B &  & 0.241 & 0.623 & 0.812 &  &  & 0.371 & 0.690 & 0.940 &  &  & 1.822 & 0.145 & 0.395 \\
Otherdeed for Otherside & A &  & 1.837 & 0.180 & 0.449 &  &  & 1.331 & 0.272 & 0.582 &  &  & 2.102 & 0.110 & 0.300 \\
 & B &  & 6.025 & \textbf{0.016} & 0.053 &  &  & 4.206 & \textbf{0.019} & 0.095 &  &  & 1.343 & 0.269 & 0.538 \\
Art Blocks Curated & A &  & 37.546 & \textbf{0.0001} & \textbf{0.001} &  &  & 11.561 & \textbf{0.0001} & \textbf{0.003} &  &  & 9.026 & \textbf{0.0001} & \textbf{0.003} \\
 & B &  & 29.050 & \textbf{0.0001} & \textbf{0.003} &  &  & 10.329 & \textbf{0.0001} & \textbf{0.003} &  &  & 5.202 & \textbf{0.0017} & \textbf{0.016} \\
Azuki & A &  & 0.372 & 0.543 & 0.740 &  &  & 1.771 & 0.1756 & 0.439 &  &  & 1.494 & 0.221 & 0.510 \\
 & B &  & 4.650 & \textbf{0.033} & 0.099 &  &  & 2.380 & 0.098 & 0.267 &  &  & 1.630 & 0.187 & 0.400 \\
CLONE X & A &  & 5.4105 & \textbf{0.021} & 0.078 &  &  & 4.864 & \textbf{0.009} & \textbf{0.045} &  &  & 3.071 & \textbf{0.031} & 0.132 \\
 & B &  & 21.746 & \textbf{0.0001} & \textbf{0.003} &  &  & 4.282 & \textbf{0.016} & 0.096 &  &  & 2.232 & 0.089 & 0.381 \\
Decentraland & A &  & 1.039 & 0.308 & 0.710 &  &  & 0.567 & 0.567 & 0.809 &  &  & 0.289 & 0.832 & 0.959 \\
 & B &  & 0.349 & 0.555 & 0.792 &  &  & 0.544 & 0.581 & 0.917 &  &  & 0.447 & 0.719 & 0.862 \\
The Sandbox & A &  & 0.321 & 0.572 & 0.746 &  &  & 0.763 & 0.469 & 0.740 &  &  & 1.438 & 0.237 & 0.507 \\
 & B &  & 0.021 & 0.884 & 0.914 &  &  & 0.700 & 0.498 & 0.878 &  &  & 0.800 & 0.496 & 0.826 \\
Moonbirds & A &  & 1.021 & 0.315 & 0.674 &  &  & 1.191 & 0.310 & 0.620 &  &  & 0.933 & 0.430 & 0.614 \\
 & B &  & 18.768 & \textbf{0.0001} & \textbf{0.003} &  &  & 9.183 & \textbf{0.0003} & \textbf{0.004} &  &  & 6.487 & \textbf{0.0007} & \textbf{0.021} \\
Doodles & A &  & 0.833 & 0.363 & 0.680 &  &  & 1.040 & 0.356 & 0.628 &  &  & 1.242 & 0.297 & 0.524 \\
 & B &  & 0.434 & 0.511 & 0.806 &  &  & 0.689 & 0.503 & 0.838 &  &  & 0.681 & 0.565 & 0.847 \\
Meebits & A &  & 0.001 & 0.979 & 0.979 &  &  & 0.036 & 0.963 & 0.996 &  &  & 0.080 & 0.970 & 0.970 \\
 & B &  & 0.140 & 0.708 & 0.786 &  &  & 0.063 & 0.938 & 0.970 &  &  & 0.058 & 0.981 & 0.981 \\
Cool Cats & A &  & 0.090 & 0.764 & 0.955 &  &  & 0.752 & 0.473 & 0.709 &  &  & 1.294 & 0.278 & 0.521 \\
 & B &  & 7.166 & \textbf{0.008} & \textbf{0.030} &  &  & 3.896 & \textbf{0.022} & 0.094 &  &  & 4.882 & \textbf{0.002} & \textbf{0.015} \\
Bored Ape Kennel Club & A &  & 0.075 & 0.784 & 0.871 &  &  & 0.351 & 0.704 & 0.812 &  &  & 0.327 & 0.805 & 0.966 \\
 & B &  & 0.201 & 0.653 & 0.783 &  &  & 0.717 & 0.489 & 0.916 &  &  & 0.964 & 0.410 & 0.768 \\
Loot (for Adventurers) & A &  & 0.961 & 0.328 & 0.656 &  &  & 0.450 & 0.638 & 0.797 &  &  & 0.166 & 0.918 & 0.949 \\
 & B &  & 0.238 & 0.626 & 0.782 &  &  & 0.067 & 0.935 & 1.000 &  &  & 0.572 & 0.634 & 0.905 \\
CryptoKitties & A &  & 2.111 & 0.146 & 0.398 &  &  & 10.377 & \textbf{0.005} & \textbf{0.030} &  &  & 16.309 & \textbf{0.0009} & \textbf{0.006} \\
 & B &  & 1.765 & 0.183 & 0.392 &  &  & 1.434 & 0.488 & 0.976 &  &  & 1.224 & 0.747 & 0.861 \\
CrypToadz & A &  & 36.987 & \textbf{0.0001} & \textbf{0.001} &  &  & 17.285 & \textbf{0.0001} & \textbf{0.003} &  &  & 11.985 & \textbf{0.0001} & \textbf{0.003} \\
 & B &  & 0.367 & 0.545 & 0.817 &  &  & 1.299 & 0.276 & 0.591 &  &  & 1.913 & 0.130 & 0.433 \\
World of Women & A &  & 4.018 & \textbf{0.046} & 0.155 &  &  & 3.518 & \textbf{0.032} & 0.106 &  &  & 2.484 & 0.063 & 0.189 \\
 & B &  & 13.267 & \textbf{0.0004} & \textbf{0.002} &  &  & 3.506 & \textbf{0.032} & 0.120 &  &  & 2.788 & \textbf{0.042} & 0.252 \\
SuperRare & A &  & 11.238 & \textbf{0.0009} & \textbf{0.006} &  &  & 3.347 & \textbf{0.036} & 0.107 &  &  & 2.706 & \textbf{0.045} & 0.15 \\
 & B &  & 0.800 & 0.371 & 0.695 &  &  & 0.396 & 0.673 & 0.961 &  &  & 0.933 & 0.424 & 0.748 \\ \midrule
Phunky Ape Yacht Club & A &  & 3.199 & 0.078 & 0.233 &  &  & 2.022 & 0.140 & 0.381 &  &  & 1.222 & 0.309 & 0.515 \\
 & B &  & 0.200 & 0.655 & 0.755 &  &  & 0.333 & 0.718 & 0.861 &  &  & 0.565 & 0.639 & 0.871 \\
PHAYC & A &  & 0.405 & 0.528 & 0.792 &  &  & 0.569 & 0.571 & 0.778 &  &  & 0.258 & 0.854 & 0.948 \\
 & B &  & 0.310 & 0.580 & 0.790 &  &  & 0.435 & 0.650 & 0.975 &  &  & 0.063 & 0.978 & 1.000 \\
CryptoPhunks & A &  & 0.678 & 0.411 & 0.684 &  &  & 0.338 & 0.713 & 0.792 &  &  & 0.229 & 0.875 & 0.937 \\
 & B &  & 0.001 & 0.997 & 0.997 &  &  & 0.066 & 0.935 & 1.000 &  &  & 0.366 & 0.777 & 0.863 \\
SameToadz & A &  & 0.077 & 0.781 & 0.901 &  &  & 0.041 & 0.959 & 1.000 &  &  & 1.343 & 0.269 & 0.538 \\
 & B &  & 0.749 & 0.390 & 0.688 &  &  & 0.356 & 0.701 & 0.876 &  &  & 0.500 & 0.683 & 0.853 \\
AIMoonbirds & A &  & 0.719 & 0.403 & 0.711 &  &  & 0.004 & 0.995 & 0.995 &  &  & 1.026 & 0.399 & 0.598 \\
 & B &  & 0.505 & 0.482 & 0.803 &  &  & 0.221 & 0.803 & 0.926 &  &  & 1.973 & 0.146 & 0.365 \\
Zukibirds & A &  & 0.090 & 0.765 & 0.917 &  &  & 0.681 & 0.574 & 0.748 &  &  & 0.878 & 0.497 & 0.677 \\
 & B &  & 2.430 & 0.131 & 0.302 &  &  & 1.711 & 0.196 & 0.452 &  &  & 2.123 & 0.122 & 0.457 \\
MoonbirdPunks & A &  & 0.411 & 0.531 & 0.758 &  &  & 1.037 & 0.386 & 0.643 &  &  & 0.357 & 0.785 & 0.981 \\
 & B &  & 3.158 & 0.097 & 0.242 &  &  & 2.387 & 0.137 & 0.342 &  &  & 0.140 & 0.933 & 0.999 \\
Undead Pastel Club & A &  & 17.414 & \textbf{0.001} & \textbf{0.006} &  &  & 8.443 & \textbf{0.0005} & \textbf{0.005} &  &  & 8.608 & \textbf{0.0001} & \textbf{0.003} \\
 & B &  & 4.418 & \textbf{0.038} & 0.103 &  &  & 3.429 & \textbf{0.037} & 0.123 &  &  & 1.938 & 0.130 & 0.433 \\
Coodles & A &  & 0.013 & 0.908 & 0.939 &  &  & 0.448 & 0.641 & 0.769 &  &  & 0.512 & 0.675 & 0.880 \\
 & B &  & 0.064 & 0.800 & 0.857 &  &  & 0.362 & 0.698 & 0.910 &  &  & 0.518 & 0.671 & 0.875 \\
Lil Baby Ape Club & A &  & 18.579 & \textbf{0.0001} & \textbf{0.001} &  &  & 7.099 & \textbf{0.001} & \textbf{0.007} &  &  & 5.556 & \textbf{0.001} & \textbf{0.006} \\
 & B &  & 8.934 & \textbf{0.003} & \textbf{0.012} &  &  & 2.374 & 0.097 & 0.291 &  &  & 1.788 & 0.153 & 0.353 \\
Lil Baby Cool Cats & A &  & 0.066 & 0.800 & 0.857 &  &  & 1.264 & 0.323 & 0.605 &  &  & 1.279 & 0.353 & 0.557 \\
 & B &  & 1.025 & 0.329 & 0.658 &  &  & 0.062 & 0.940 & 0.940 &  &  & 0.770 & 0.546 & 0.862 \\ \bottomrule
\end{tabular}
}
\label{Granger_causality_result}
\end{table}

\section{Granger causality Test on Tweets number and NFT Price}
Utilizing the tweets and transactions collected, we investigate the causal relationship between the social network activeness and the NFT average prices temporally, then compare the results for the original projects and copycat projects.
We choose to perform a Granger causality test given the number of tweets and the average transaction values (price of the traded NFTs) within consecutive timeframes.
Granger causality presents the ability of the (lagged) time series $X$ to help predict another time series $Y$ from the information that time series $X$ contains.
Granger causality is defined in \cite{granger1981some} as series $X$ does not cause the series $Y$ if: 

\begin{equation*}
    Y_{t+1} \indep \mathcal{I}(t) | \mathcal{I}\__X(t)
\end{equation*}

Double tack up \indep represents conditional independence of random variables and $\mathcal{I}(t)$, $\mathcal{I}\__X(t)$ denote the information in time $t$, and the information in time $t$ without the information of $X$.
The series can be $X$: the number of tweets and $Y$: the average transaction values or inversely.
Since these are multiple tests existing for comparison, i.e., each NFT project is a different test, the hypothesis tests should be conducted with a method to reduce the chances of false-positive results.
Here we use Benjamini-Hochberg correction \cite{benjamini1995controlling}, but not Bonferroni correction \cite{weisstein2004bonferroni} as it is conservative for possible rejection of false null hypothesis.
The corrected $P$ value is defined as below, where $k$ is the ascending order of the $P$ values among all tests, and $m$ is the number of tests.

\begin{equation*}
    Corr\, P \, = P \times \frac{k}{m}
\end{equation*}

We set one individual timeframe (tf) to be of length 3 days.
Within a timeframe, the total number of tweets and the average prices of the traded NFTs are calculated (with the transfer transactions filtered out).
We present the Granger causality test results in Table. \ref{Granger_causality_result}.  
Note that all the projects come with 2 types of null hypotheses (A and B) and 3 choices of lags: 1 timeframe (3 days), 2 timeframes (6 days), and 3 timeframes (9 days).
As for the original projects, 10/12 out of 19 original projects show corrected/uncorrected null hypothesis rejection, and 6 show A null hypothesis rejection indicating that the number of tweets contains information that helps predict the average traded price of NFTs for those projects.
7 original projects show B null hypothesis meaning that the NFT price has a significant impact on the number of tweets, corresponding to the scenarios where the falling prices cause fewer tweets or the rising prices boost more tweets.
By contrast, the fake or copycat projects show weak Granger causality.
Only 2 copycat projects considered derivatives: Lil Baby Ape Club and Undead Pastel Club show positive Granger causality between the number of tweets and the NFT prices.

\begin{figure}[!htbp]
	\centering
	\includegraphics[width=0.471\textwidth, scale=1]{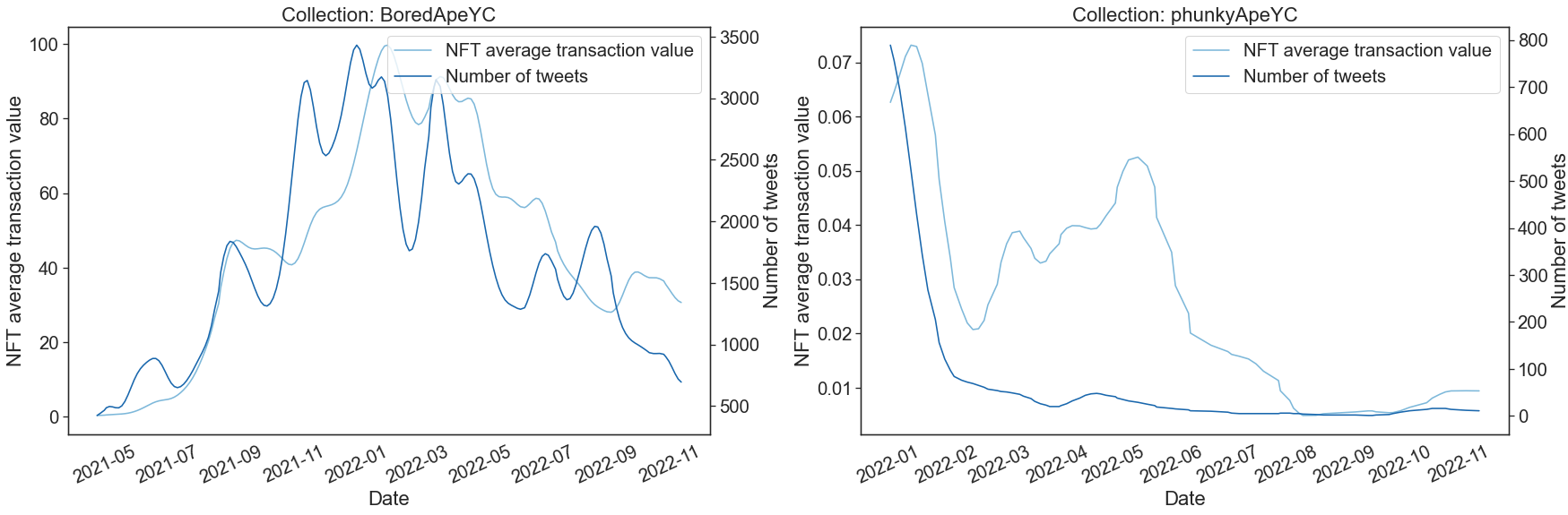} 
	\caption{Example of number of tweets and NFT average transaction value for 
	an authentic project: BAYC and a copycat project: PhunkyAYC.}
	\label{fig:compare_authentic_projects}
\end{figure}
The reason for the causal relationship being insufficient on the copycat projects as we observe: for copycat projects, no sufficient corresponding proportion of increased tweets appeared when the price NFT price increased, or no attention is given to some copycat projects after a particular time therefore almost no new tweets anymore.
Figure. \ref{fig:compare_authentic_projects} shows an example of an authentic project compared to a copycat project.
Granger causality is marginal predictability indicating whether one time series is useful in forecasting another.
This marginal predictability exists when similar patterns of peaks and plateaus in a time series data happened at the same time or prior (add lags for offsetting the time gap) to another time series data.
Based on the observation that many plateaus exist in the number of tweets for the copycat projects, which may cause insufficient causality as the price at that time may still fluctuate.
We perform a plateau detection by taking the derivatives of chunked points and counting the percentage of points with a max slope of .001. The results are shown in Table. \ref{tab:percent_plateau_table}, which supports our observation and reason that hypothesizes the insufficient causal relationship.
Our causality tests do have limits: lags is maximum 3 timeframes (9 days), which may lead to false negative for both authentic and copycat projects.
However, under the existing experiments, we do find marked evidence for larger part of the authentic projects having positive impact relationships between the number of tweets and the prices or reversely.

\begin{table}[t!]
\huge
\caption{Percentage of plateaus for the number of tweets and NFT average transaction values}
\centering
\resizebox{0.41\textwidth}{!}{%
\begin{tabular}{lcc}
 & \multicolumn{1}{l}{Number of tweets} & \multicolumn{1}{l}{NFT average transaction value} \\ \hline
Authentic projects & 0.1191 & 0.1129 \\
Copycat projects & 0.5055 & 0.1117
\end{tabular}
}
\label{tab:percent_plateau_table}
\end{table}

\begin{figure*}[!htbp]
	\centering
	\includegraphics[width=0.9997\textwidth, scale=1]{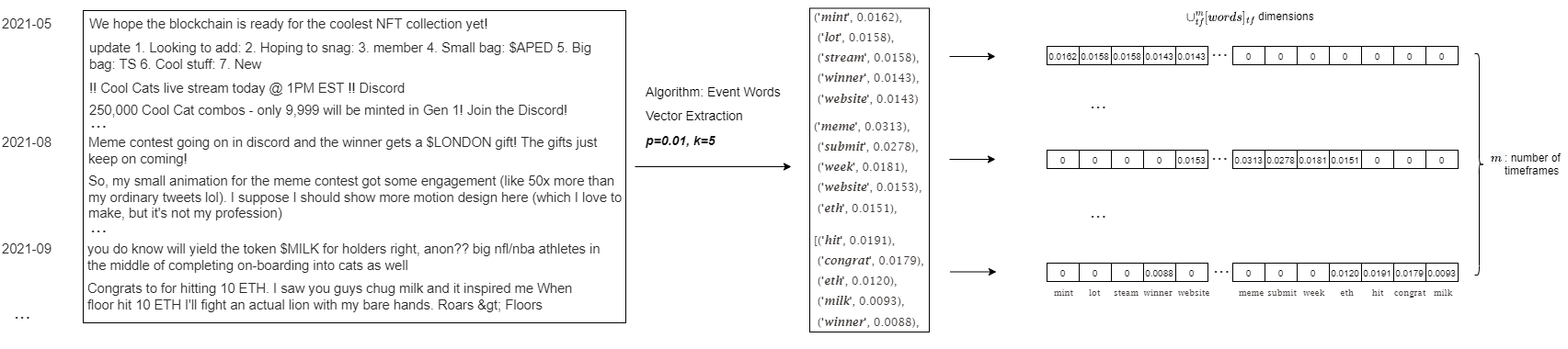} 
	\caption{Example of feature words vector extraction process (Tweets from Cool Cat NFT).}
	\label{fig:event_word_process}
\end{figure*}

\section{NFT price moves Prediction}
After we see the evidence of the positive impact of the number of tweets on the NFT price, we then explore the content of the tweets and the features behind the content, which will potentially reveal the hidden factors that drive up the NFT prices.
Our inspiration is based on the observation that the growth of an NFT project is accompanied by a series of official project events such as prior-mint promotion, release for sale, airdrop, derivative NFT announcement, DeFi or GameFi connection; and community-based events such as the interaction with influencers or celebrities, the creation of memes, the engagement of various online or offline activities.
In this section, we propose a method to extract important words with TFIDF scores from the tweets, which we use as the features to investigate the relation between these features and the NFT price moves.
Then we set up a price moves prediction task given the features and perform an analysis of the results.
%


\subsection{Feature Words Extraction on Tweets}

In social media communities, languages consist of a wide range of expressions about the events and the related behaviors. 
A method to extract the words describing the events from the tweets without prior knowledge of any events will help in building the word features.
We first divide all the tweets into groups, where each group contains the tweets in a timeframe of 3 days so that we have a list of groups of tweets ordered by date.

\algnewcommand\algorithmicforeach{\textbf{for each}}
\algdef{S}[FOR]{ForEach}[1]{\algorithmicforeach\ #1\ \algorithmicdo}

\begin{algorithm}
\caption{Feature Words Vector Extraction}
\label{algo: word_vector}
\small

\begin{flushleft}
\textbf{Data:} Tweets in string for a collection \\
\textbf{Result:} Event word vector for each timeframe
\end{flushleft}

\begin{algorithmic}[1]
\State ${tweets} \leftarrow removeLinksTagsEmoji()$
\State $tweets\_in\_tfs \leftarrow splitInTimeframes(tweets, n)$
\State ${tweets\_in\_tfs} \leftarrow removeFramesNoTransaction()$
\Procedure{GetWords}{$tweets\_in\_tfs$, $p$}
\ForEach {$tweet$ in $tweets\_in\_tfs$}
    \State ${[word]} \leftarrow ExtractWithPOS(tweet) $
    \State \texttt{ \quad \quad \quad \quad If $word.pos$ is VERB|NOUN}
\EndFor
\ForEach {$word$ in a timeframe $tf$}
    \State $tfidf \leftarrow CalTFIDF(word, p) $
\EndFor
\EndProcedure
\For {$[tfidf]$ of words in a timeframe $tf$}
    \State ${vector} \leftarrow FeatureWordVectorize( [tfidf], k ) $
\EndFor

\end{algorithmic}
\end{algorithm}

We perform a feature words vector extraction method described in Algorithm. \ref{algo: word_vector} to extract the features from the tweets of a timeframe, which is represented by a vector of the importance score of relevant words.
We first extract all nouns and verbs for each tweet using the tool Spacy \cite{Spacy_POS} with the English Part-of-speech (POS) tagging model.
After obtaining the list of nouns and verbs, we use a term frequency-inverse document frequency (TF-IDF) method to have a list of the retrieved important words and the weights of importance for the content of that timeframe.
%
%
%
Here we treat the extracted nouns and verbs from the tweets in a timeframe as a document that describes the topics or events for this timeframe.
We consider the words extracted in a timeframe can likely represent the most widely used keywords that appeared mostly in that timeframe compared to other timeframes.
For example, in some early timeframes, \textit{mint} is of a large TF, and \textit{mint} will not occur with a high frequency in all timeframes. So some timeframes where the word \textit{mint} occurs with a large TF-IDF are most likely to be the time when some mint events happened.
We add a parameter \textit{p} in line 4 and 10 in our method to be the minimum frequency set as $.01 \cdot |W|$ for a word to be considered as being contained in a timeframe that contains a set of words $W$.
The minimum frequency is for preventing the words that are mentioned in all the timeframes but within only a few timeframes frequently mentioned from being with a small IDF score.
%
%
In contrast, IDF weights down the words that frequently appear in most timeframes such as \textit{love}, \textit{cool}, \textit{join}, etc., in the output TF-IDF score.
After performing the TF-IDF method, we will make the top $k$ TF-IDF values of words in a timeframe a vector, where the vector dimension is the union of the $k$ words of all $m$ timeframes: $\cup_{tf}^{m} [words]_{tf}$.
We set $k$ default to 60 in the following experiment.
The vectors will be used as input features for modeling the relation between the keywords and NFT price moves.
An example of the process of the feature words vector extraction is demonstrated in Figure. \ref{fig:event_word_process}, where we use Cool Cat NFT as an example.
The words extracted in those three timeframes, after we confirm by searching news, correspond to the events of the announcement of mint in Jun 2021, the Cool Cat meme competition in Aug 2021, and the first time cool cat floor price hit 10 ETH in Sept 2021.
However, we don't have the ground truth labels for the events of the NFT projects we collected to perform the validation through mapping the retrieved words to each individual event. 
Nevertheless, the words are features inherently describing the events or focused discussion topics that may be unique to that timeframe.
We run Algorithm. \ref{algo: word_vector} on the tweets of all 19  authentic NFT collections.
%
%
%
%
%
The average number of timeframes among all the authentic projects is 188 timeframes, and the average number of the total extracted words is 1781 words.
%


\subsection{Normalized Price Regression}
\label{sec:price_prediction}
With the extracted words vectors as the input features, we can further investigate the predictability of NFT price moves using machine learning regression models.
The prediction tasks will provide understanding in terms of both the predictability of NFT price given social media keywords vectors, and the question of which features contribute to the prediction positively or negatively more than other features for further analysis.
Since each feature (dimension) of the feature vector is a weight value for a particular word,  we intend to discover the insights of the NFT social media communities and NFT prices by extracting the characteristics of these words.

\begin{table*}[!htbp]
\huge
\caption{Evaluation results of the Markov normalized price prediction.}\label{tab:total_table}
\centering
\resizebox{0.98\textwidth}{!}{
\begin{tabular}{lccclccclccclcccc}
\hline
Project (Collection) & Date Range & Timeframes & Features &  & \multicolumn{3}{c}{Acc} &  & \multicolumn{3}{c}{F1} &  & \multicolumn{3}{c}{MAE} & \multicolumn{1}{l}{} \\ \cline{6-8} \cline{10-12} \cline{14-16}
 & \multicolumn{1}{l}{} & \multicolumn{1}{l}{} & \multicolumn{1}{l}{(Words)} &  &  \multicolumn{1}{l}{ \hspace{1em} SVM \hspace{1em}\hspace{1em}}   &   \multicolumn{1}{l}{\hspace{1em} MLP \hspace{1em}\hspace{1em}}   & \multicolumn{1}{l}{Transformer} &  & \multicolumn{1}{l}{ \hspace{1em} SVM \hspace{1em}\hspace{1em}} & \multicolumn{1}{l}{\hspace{1em} MLP \hspace{1em}\hspace{1em}} & \multicolumn{1}{l}{Transformer} &  & \multicolumn{1}{l}{\hspace{1em} SVM \hspace{1em}\hspace{1em}} & \multicolumn{1}{l}{\hspace{1em} MLP \hspace{1em}\hspace{1em}} & \multicolumn{1}{l}{Transformer} & \multicolumn{1}{l}{} \\ \hline
CryptoPunks & 2018-01-02 to 2022-11-14 & 221 & 1633 &  & 0.545 & 0.628 & 0.667 &  & 0.706 & 0.694 & 0.648 &  & 0.277 & 0.291 & 0.233 &  \\
Bored Ape Yacht Club & 2021-04-24 to 2022-11-15 & 191 & 1717 &  & 0.421 & 0.622 & 0.754 &  & 0.577 & 0.631 & 0.697 &  & 0.256 & 0.234 & 0.23 &  \\
Mutant Ape Yacht Club & 2021-08-29 to 2022-11-14 & 148 & 1593 &  & 0.413 & 0.667 & 0.667 &  & 0.585 & 0.564 & 0.565 &  & 0.181 & 0.170 & 0.169 &  \\
Otherdeed for Otherside & 2022-05-01 to 2022-11-13 & 65 & 1164 &  & 0.538 & 0.622 & 0.754 &  & 0.571 & 0.631 & 0.697 &  & 0.294 & 0.234 & 0.23 &  \\
Art Blocks Curated & 2020-12-12 to 2022-11-15 & 235 & 2336 &  & 0.617 & 0.652 & 0.574 &  & 0.571 & 0.625 & 0.432 &  & 0.411 & 0.412 & 0.417 &  \\
Azuki & 2022-01-12 to 2022-11-15 & 103 & 1585 &  & 0.600 & 0.717 & 0.733 &  & 0.714 & 0.730 & 0.765 &  & 0.226 & 0.215 & 0.223 &  \\
CLONE X & 2021-12-13 to 2022-11-11 & 115 & 893 &  & 0.571 & 0.650 & 0.667 &  & 0.526 & 0.550 & 0.556 &  & 0.154 & 0.156 & 0.153 &  \\
Decentraland & 2019-01-22 to 2022-11-15 & 467 & 2431 &  & 0.500 & 0.613 & 0.62 &  & 0.667 & 0.627 & 0.642 &  & 0.155 & 0.167 & 0.156 &  \\
The Sandbox & 2022-01-30 to 2022-11-13 & 91 & 1397 &  & 0.722 & 0.759 & 0.777 &  & 0.615 & 0.705 & 0.702 &  & 0.257 & 0.254 & 0.256 &  \\
Moonbirds & 2022-04-16 to 2022-11-13 & 74 & 1004 &  & 0.428 & 0.666 & 0.690 &  & 0.200 & 0.563 & 0.602 &  & 0.137 & 0.128 & 0.122 &  \\
Doodles & 2021-10-17 to 2022-11-15 & 132 & 1509 &  & 0.461 & 0.640 & 0.666 &  & 0.632 & 0.592 & 0.609 &  & 0.246 & 0.233 & 0.238 &  \\
Meebits & 2022-03-21 to 2022-11-13 & 80 & 1486 &  & 0.500 & 0.625 & 0.604 &  & 0.556 & 0.591 & 0.577 &  & 0.468 & 0.462 & 0.47 &  \\
Cool Cats & 2021-07-02 to 2022-11-15 & 168 & 1615 &  & 0.545 & 0.596 & 0.606 &  & 0.681 & 0.570 & 0.626 &  & 0.217 & 0.226 & 0.205 &  \\
Bored Ape Kennel Club & 2021-06-19 to 2022-11-15 & 172 & 1653 &  & 0.705 & 0.617 & 0.667 &  & 0.688 & 0.580 & 0.610 &  & 0.144 & 0.167 & 0.141 &  \\
Loot (for Adventurers) & 2021-08-28 to 2022-11-15 & 144 & 2316 &  & 0.678 & 0.702 & 0.726 &  & 0.69 & 0.614 & 0.684 &  & 0.325 & 0.330 & 0.338 &  \\
CryptoKitties & 2018-09-18 to 2022-11-13 & 506 & 3379 &  & 0.485 & 0.590 & 0.597 &  & 0.409 & 0.415 & 0.365 &  & 0.714 & 0.706 & 0.693 &  \\
CrypToadz & 2021-09-09 to 2022-11-13 & 144 & 2133 &  & 0.535 & 0.607 & 0.750 &  & 0.480 & 0.466 & 0.682 &  & 0.134 & 0.153 & 0.127 &  \\
World of Women & 2021-07-28 to 2022-11-14 & 158 & 1553 &  & 0.548 & 0.699 & 0.709 &  & 0.500 & 0.489 & 0.482 &  & 0.186 & 0.191 & 0.186 &  \\
SuperRare & 2019-09-12 to 2022-11-14 & 375 & 2443 &  & 0.480 & 0.613 & 0.573 &  & 0.571 & 0.524 & 0.474 &  & 0.593 & 0.583 & 0.579 &  \\ \hline
\end{tabular}
}
\label{prediction_result}
\end{table*}

\subsubsection{Method}

The model we use include: 
\noindent \textit{multilayer perceptron (MLP)}: two dense layers of 64, 256 units with batch normalization layer in between and the last output layer of 1 unit for the regression output. 
\noindent \textit{SVM}: Support Vector Machine regressor for regression with the regularization parameter C between 0.1 to 10.0 and pick up the one with the best performance. 
\noindent \textit{Transformer}: Light Transformer with two attention heads and small 32 dimensions for embedding size and hidden layer size, adopted from the same setting in \cite{apoorv_nandan_notes_2020} for IMDB sentiment classification.
We split all the timeframes of a project into the latest 20\% for testing and the rest 80\% for training by the date time.
The input features are the word vectors for one timeframe of the project, and the ground truth is the normalized NFT average price of the next timeframe.
The reason for adding the lag to the words features is to predict the price of a future timeframe given the words features from a previous timeframe.
For example, for BAYC NFT, in the timeframe of July 1, 2021, the average trading price is 3.37 ETH while in the timeframe of Feb 25, 2022, the average trading price is 90.32 ETH, which demostrates a magnitude difference in a price growing trend. 
The trend is created through market behaviors of the whole NFT development, therefore, predicting the raw price given the feature words vector is not much hope work for finding the relation of certain events described by words and NFT price.
Instead, we explore predicting the price changing proportion, i.e., the ratio of current price and the average price of several timestamps prior to the current timestamp.
Inspired by some early works using Markov assumption for NLP tasks \cite{wolkowicz2013evaluation} where the assumption is to calculate the probability of a symbol to occur only depends on its previous $n$ symbols, we use the normalization:
\begin{equation*}
  y_{i}^\prime  = \frac{y_{i}}{\sum_{k=i-n}^{i}y_{k}/n}
\end{equation*}
$y_{i}$ is the raw price at the timeframe $i$, and the Markov normalized price $y_{i}^\prime$ will calculated by being divided by the average of its previous $n$ raw prices.
The first $n$ timeframes will be dropped since have no previous timeframes.
After the normalization, for example, with an $n=3$, a normalized value of 1.12 means the average NFT price of the current timeframe is 12\% larger than the average price of its previous timeframes of length 3.
In the meanwhile, we use a mean absolute error (MAE) loss with a penalizer $\delta_{i}$ ($\delta_{i}=1$ if the ground truth $y_{i}$ and prediction $\hat{y}_i$ are both >1 or both <1. otherwise $\delta_{i}=1.5$) for wrong price moves predictions as the equation below shows.

\begin{equation*}
    loss = \delta_{i} * \frac{1}{m} \sum_{i=1}^{m} |y_{i}^\prime - \hat{y}_i|
\end{equation*}

\begin{figure*}[!htbp]
	\centering
	\includegraphics[width=0.876\textwidth, scale=1]{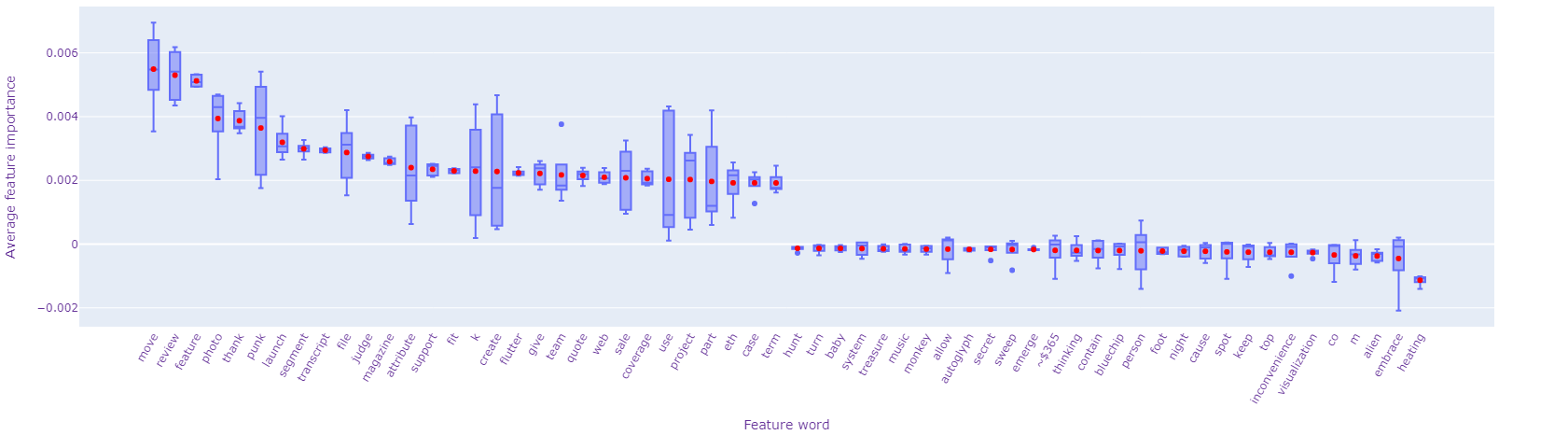}
	\caption{Example of Mean Decrease Accuracy (MDA) feature importance on one of the collections: CryptoPunks. Mean and variance are made from 5 repeat times of permuting a feature. Top 30 words with the largest mean importance values are followed by the 30 words with the smallest mean importance values from left to right.}
	\label{fig:importance_keywords}
\end{figure*}

\subsubsection{Results}

We evaluate our model using both the regression metric MAE and the classification metrics (accuracy and F1 score) with three runs of execution and take the average.
The classification metrics are used since the metrics help in perceiving the correctness of the prediction of movement and our regression results can be easily reformed to a binary classification results by converting the prediction and the ground truth to 1 or 0 representing the price moves up or down ($y_{i}^\prime := 1 \text{ if } y_{i}^\prime > 1 \text{ else 0}, \text{ the same for } \hat{y}_i$).
The prediction results shown in Table. \ref{prediction_result} present a better than a random baseline of .5 accuracy price moves predictive performance for all the collections using the MLP and the Transformer.
The MLP obtains an average accuracy of 64.6\% among all the collections, and the Transformer obtains 67.3\%, and the Transformer achieves a better accuracy compared to the MLP in 16 out of 19 projects.
Since we use a Markov window of length 3 to confine the normalization calculated on a short period prior to the current price, the predictability reflects the words as the predictors of a relatively quick NFT price change within 3 timeframes (9 days) representing small liquidity.
It is important to note that the words used for discussion on social media about the NFTs are inherently not decisive predictors for price moves since the price change is accompanied by heterogeneous market behaviors.
Nevertheless, the prediction tasks show certain predictability with the words vectors as predictors.
The results provide the rationality of further investigation of which words (features) take effect on the prediction of price moves more compared to others, which will provide insights into NFT social media contents accompanied with the price moves.

\subsubsection{Analysis}

We use the Mean Decrease Accuracy (MDA) method on the models for each collection to compute the importance of the features (words).
The method calculates the importance by measuring how much the validation metric degrades when a single feature value is randomly shuffled \cite{breiman2001random}: the feature importance score is defined as:
\begin{equation*}
    score_i = s - \frac{1}{k}\sum_{j=1}^k(s_{i,j})
\end{equation*}
For each feature $i$ in vectors, randomly shuffle column $i$ to generate an edited dataset and repeat $k$ (we use $k$=5) times, so we get the edited datasets $\{ \tilde{D}_{i,j}\}, j \in \{0,1,..,k\}$ and compute the score $s_{i,j}$ using the predictive model to fit each edited dataset $\tilde{D}_{i,j}$. $s$ is the score obtained by fitting the predictive model on the original dataset $D$.
We use the model with the best prediction performance as the predictive model. As a result, we obtain a list of importance scores for each of the feature words.
We trim out the words with the top $n$ highest positive scores followed by the lowest $n$ negative importance scores.
Figure. \ref{fig:importance_keywords} illustrates an example of the top 30 positive and negative keywords of features importance.
The positive values of importance indicate that the words make positive contributions to the normalized price prediction task, which means these words are good predictors for the price moves.
Further, we attempt to analyze the characteristics of the MDA top positive words compared to negative words to provide insight into which characters of words can help monitor the price moves. 
The characteristics include categories and sentiment.
We observe general market-related words, and NFT event-related words frequently appear in the top 50 positive words for all of the projects.
To quantify the categories of the words, we adopt a zero-shot classification pipeline \cite{yin2019benchmarking} with a BART language comprehension model \cite{lewis2020bart} pretrained on MultiNLI (MNLI) dataset \cite{N18-1101} for categorizing the top keywords.
The whole zero-shot classification setup is to utilize the model pretrained on natural language inference (NLI) dataset to classify the keywords as the NLI premise and to construct a hypothesis from customized candidate labels, where the hypothesis is "This \{word\} is about \{candidate label\}", which is simple yet surprisingly effective evaluated on social media topic and emotion benchmarks in \cite{yin2019benchmarking}.
We set the candidate labels \textit{market}, \textit{non-fungible tokens}, and \textit{community} representing market-related words, NFT event-related words, and other words in social media communities with some example words shown in Table. \ref{keywords_categrization_table}.
\begin{table}[!htbp]
\huge
\caption{Example of inferenced scores from zero-shot text classifier using BART model pretained on MNLI dataset}\label{tab:total_table}
\centering
\resizebox{0.41\textwidth}{!}{%
\begin{tabular}{lcccccclll}
\multicolumn{1}{c}{} & feature & airdrop & price & profile & punk & family & buy & love & thank \\ \hline
market & 0.28 & 0.35 & \textbf{0.63} & 0.33 & 0.37 & 0.01 & \textbf{0.62} & 0.21 & 0.31 \\
non-fungible tokens & 0.27 & \textbf{0.37} & 0.16 & 0.19 & \textbf{0.54} & 0.03 & 0.13 & 0.17 & 0.21 \\
community & \textbf{0.43} & 0.26 & 0.20 & \textbf{0.47} & 0.08 & \textbf{0.96} & 0.24 & \textbf{0.61} & \textbf{0.48}
\end{tabular}
}
\label{keywords_categrization_table}
\end{table}

\begin{figure}[!htbp]
	\centering
	\includegraphics[width=0.317\textwidth, scale=1]{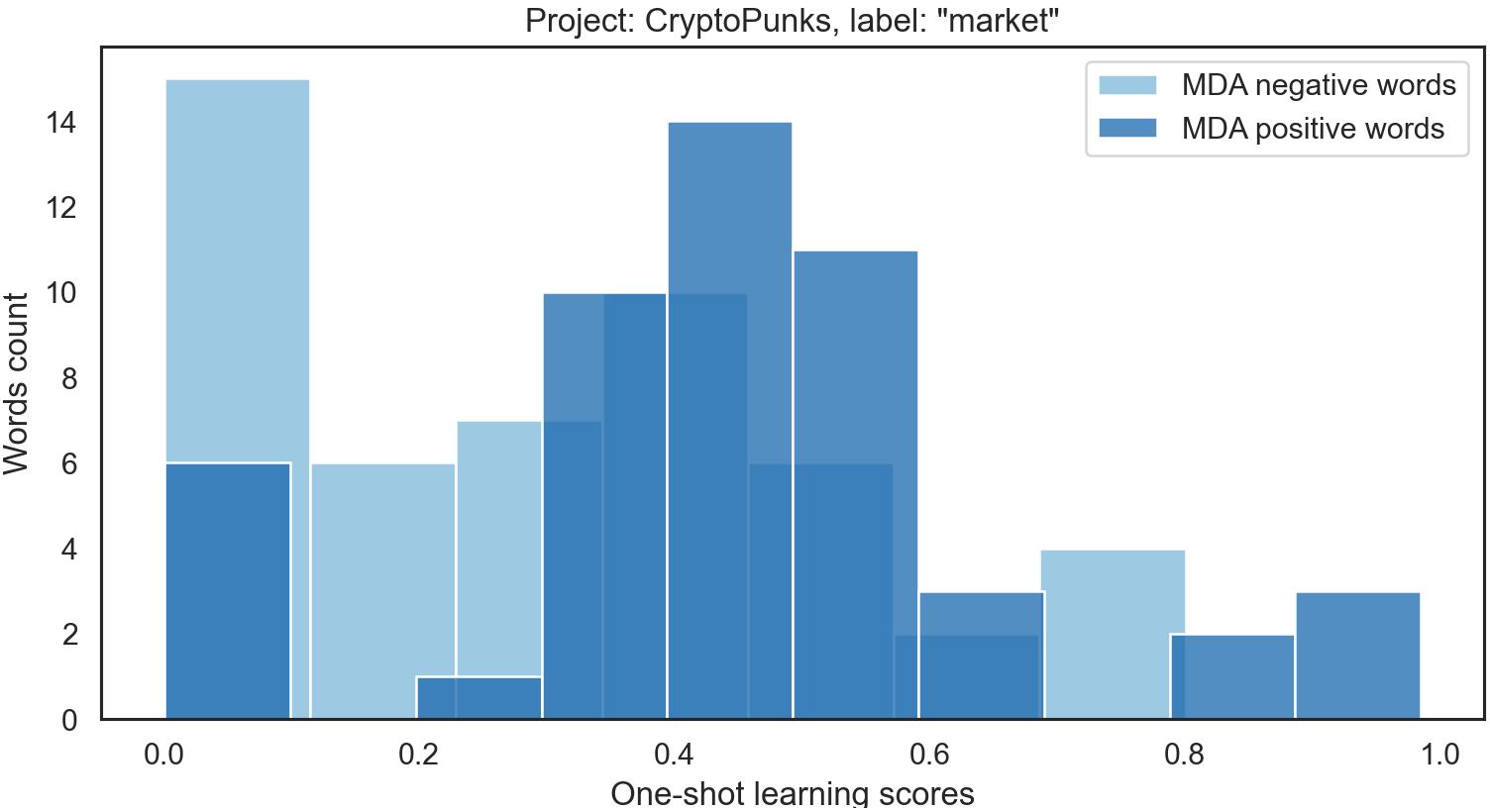} 
	\caption{Example of the word count against the inferenced score ranges of label \textit{market} for the project CryptoPunks.}
	\label{fig:inferenced_score_ranges}
\end{figure}

\begin{figure*}[!htbp]
	\centering
	\includegraphics[width=0.978\textwidth, scale=1]{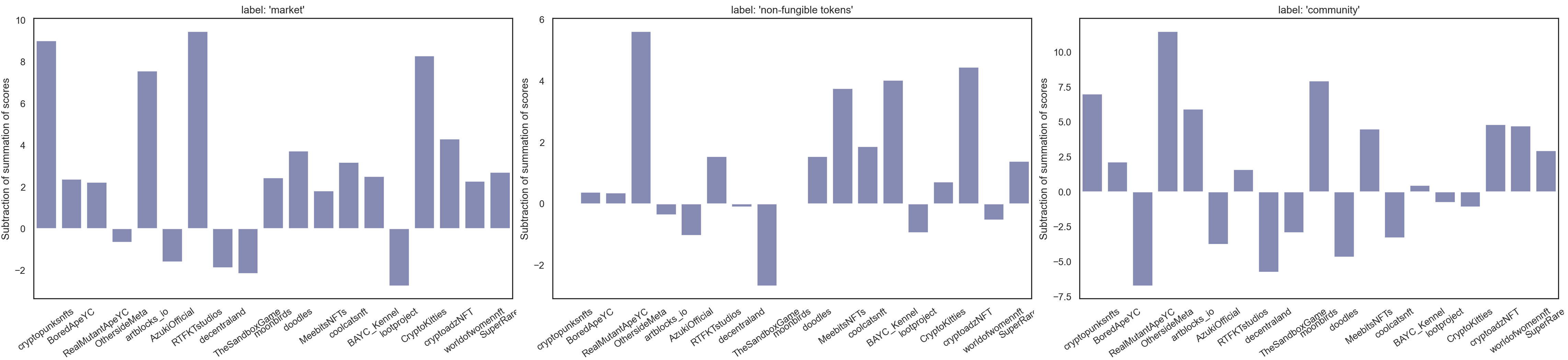} 
	\caption{The category scores differences between MDA top positive words and negative words of the candidate labels: \textit{market} (left), \textit{non-fungible tokens} (middle), and \textit{community} (right) for all projects.}
	\label{fig:score_difference_market_non-fungible-token_community}
\end{figure*}

We apply the zero-shot classification to all the top 50 positive keywords and 50 negative keywords for each project.
Figure. \ref{fig:inferenced_score_ranges} gives an example of the distribution of the number of words in inferenced score ranges, where we observe that the sum of scores for MDA positive words will be larger than the negative words taking a threshold of > .5, which evidences that more market-related words are in the positive words.
To visualize the difference of the zero-shot inferenced scores for each candidate label for MDA top positive keywords compared to negative scores, we sum up the scores of the positive words with a threshold > .5, then do the same for the negative words.   
We take the subtraction between the sum of scores of MDA positive words and negative words for each candidate label, and the results are presented in Figure. \ref{fig:score_difference_market_non-fungible-token_community}.
\begin{figure}[!htbp]
	\centering
	\includegraphics[width=0.341\textwidth, scale=1]{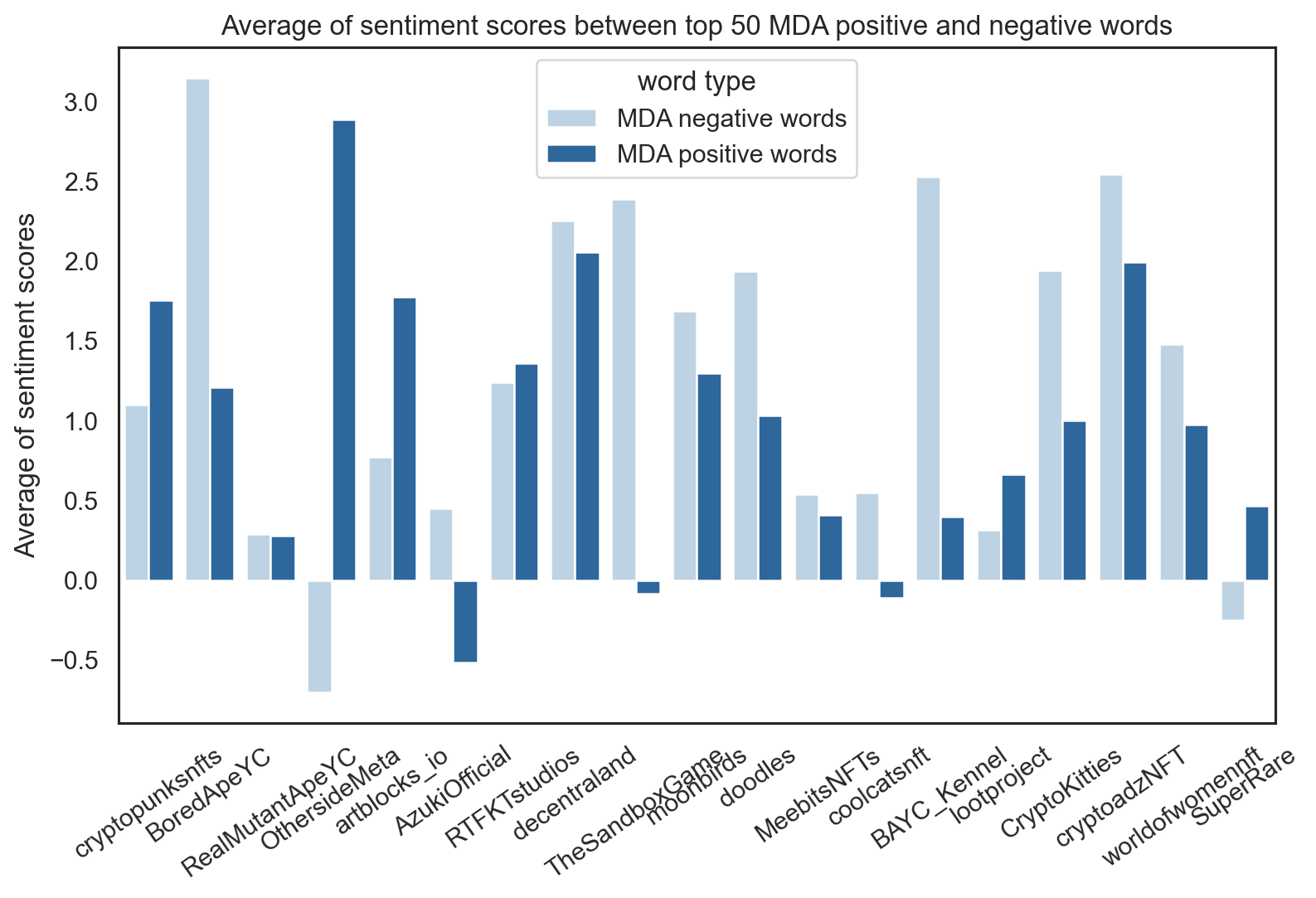}
	\caption{Total sentiment scores for all projects comparing top 50 MDA positive words with negative words.}
	\label{fig:keywords_sentiment}
\end{figure}
The label \textit{market} shows markedly positive subtraction values between MDA positive and negative words, with 5 out of 19 projects having negative results, with an average value of +4.42 for the projects with a positive subtraction compared to -1.80 for the projects with a negative subtraction.
The label \textit{non-fungible tokens} ends up with 6 out of 19 projects having negative results and +2.33 for all positive subtractions and -0.71 average values for all negative subtractions, and the label \textit{community} has 8 out of 19 projects showing negative subtraction results with +4.87 and -3.61 average values, which shows less positive subtraction values.
The results demonstrate that market-related words and NFT event-related words appear more in MDA top positive keywords, which have more contributions in predicting price moves compared to other words.
In addition to words categorized labels, sentiment has been a widely studied factor in stock price prediction \cite{balaji2017survey}.
We intend to investigate the sentiment of the top MDA positive and negative keywords.
We applied a pretrained NLP sentiment model VADER \cite{hutto2014vader} for sentiment analysis of social media text to the top MDA positive and negative keywords.
The output score from VADER for each word is -4 for most negative, +4 for most positive sentiment. 
The results of the average scores of the top 50 MDA keywords are presented in Figure. \ref{fig:keywords_sentiment}.
More than half of the projects do not show readily apparent sentiment difference comparing the positive words with the negative words except for Bored Ape Yacht Club, The Sandbox, and Bored Ape Kennel Club with more positive sentiment in MDA negative words, Otherdeed for Otherside and Art Blocks Curated with more positive sentiment in MDA positive words (focus on the projects with sentiment score difference >1).
Overall, the sentiment experiment illustrates that most words are mostly neural or positive words for both the MDA positive and negative words, and no broad sentiment difference among the majority of the projects.
Compared to sentiment, the categories: the words are market-related or event-related better correspond to the importance as price predictors

\vspace{-0.27cm}

\subsubsection{Discussion}

%
%
We observe from the MDA results that some general market-related words frequently appear in the positive words for almost all of the projects.
These general market-related words include \textit{buy}, \textit{owner}, \textit{floor}, \textit{price}, \textit{wallet}, \textit{holder}, \textit{market}, \textit{sale}, \textit{sell}, \textit{money}, \textit{own}, \textit{offer}, \textit{transaction}, \textit{volume} \textit{eth}, \textit{flip},  \textit{earn}, etc.
%
%
We suspect that the word \textit{owner} or \textit{holder} is associated with NFT whales or celebrities who bought or have been owning some NFTs mentioned in the social media or press. 
One example in history is: In May 2021, a big whale NFT collector "Pranksy" bought a large number of BAYCs and the news spread rapidly on social media, which led to a speedy sold out for the entire BAYC collections \cite{omr}.
The influence of the whales or celebrities is inevitably one of the most probable NFT price-rising reasons.
%
%
In the meanwhile, the word \textit{eth} also frequently appears, which evidences the research conducted mentioning the positive correlation between Ethereum price and NFT sales \cite{ante2022non}.
Besides, the general NFT event-related words such as  \textit{mint}, \textit{airdrop}, \textit{avatar}, \textit{pfp},
\textit{derivative}, \textit{roadmap}, \textit{founder}, also comprise a proportion of positive words.
We notice some emotional keywords such as FOMO (Fear of missing out), and FUD (fear, uncertainty, and doubt) appeared in MDA positive keywords, which may be related to an emergency motion due to a sudden event such as the FTX's collapse in Nov, 2022 \cite{reiff_2022} causing NFT price dropped, while some other keywords about concerns such as \textit{scammer}, \textit{delist}, \textit{suit}, \textit{fail}, \textit{lose}, \textit{miss} appear in MDA negative words meaning those keywords do not help in predicting price moves.
These findings are worth further work on analyzing the impact of sudden events and emerging motion, not the whole sentiment of keywords, on NFT and cryptocurrency market, as our study demonstrates that the MDA top and least important keywords show more neural or positive sentiment.
%
%
%
%
Another situation is the events that are not related to market such as the word \textit{vibemas} in MDA negative words of CrypToadz, is a slogan as 'Merry !vibemas' or originally '!vibe' which is a trigger word in CrypToadz's Discord for the bot to print out some turtle stickers.
This word \textit{vibemas} is an example of some words that may frequently pop up in a specific period, which will not benefit the prediction since the market behavior may not be correlated with sudden social media hilarity possibly because of a holiday celebration.
%



\section{Conclusion and Future Work}
This paper contributes to exploring the relationship between the NFT social media communities and the NFT price in terms of the tweet number and the content of the tweets.
We first present positive results of a Granger causality test between the number of tweets and the prices time series for more than half of the authentic projects, compared to insufficient Granger causality for most of the copycat projects.
Later we perform feature words extraction and apply the machine learning regression models for predicting a Markov normalized price given the extracted word vectors.
The results show a certain level of predictability for the normalized price.
Lastly we analyze the feature importance and summarize the findings of characteristics behind the words.
%

%
%
%
%
%
%
%
%



\bibliographystyle{ACM-Reference-Format}
\bibliography{sample-base}










\end{document}